\documentclass[twocolumn,nofootinbib,superscriptaddress,prl]{revtex4-1}
\usepackage{latexsym,epsfig,amssymb, amsmath,nicefrac}

\usepackage{color}
\usepackage[makeroom]{cancel}
  \usepackage{latexsym}
  \usepackage{epsf}
  \usepackage{amssymb}
  \usepackage{graphicx}
  \usepackage{amsmath}
  \usepackage{textcomp}
  \usepackage{amsmath,amssymb,amsthm}
  \usepackage{verbatim}
  \usepackage{hyperref}

\def\tb0{\tilde{\beta}_0}
{\def\b0{\beta_0}

\def\bi{\begin{itemize}}
\def\ei{\end{itemize}}
\def\be{\begin{equation}}
\def\ee{\end{equation}}
\newcommand{\bea}{\begin{eqnarray}}
\newcommand{\eea}{\end{eqnarray}}

\def\Kahler{K\"{a}hler~}
\def\K{K{\"a}hler}
\def\Mobius{M\"{o}bius~}

\newcommand{\eqn}[1]{(\ref{#1})}

              \newcommand{\rf}[1]{(\ref{#1})}

\parskip 7pt 

\begin{document}

\vspace{1cm}

\title{The Hyperbolic Geometry of Cosmological Attractors}

\author{John Joseph M. Carrasco}
\email{jjmc@stanford.edu}
\author{Renata Kallosh}
\email{kallosh@stanford.edu}
\author{Andrei Linde}
\email{alinde@stanford.edu}
\affiliation{Department of Physics and SITP, Stanford University, \\ 
Stanford, California 94305 USA}
\author{Diederik Roest}
\email{d.roest@rug.nl}
\affiliation{Van Swinderen Institute for Particle Physics and Gravity, University of Groningen, \\ Nijenborgh 4, 9747 AG Groningen, The Netherlands}

\begin{abstract}

Cosmological $\alpha$-attractors give a natural explanation for the spectral index $n_s$ of inflation as measured by Planck while predicting a range for the tensor-to-scalar ratio $r$, consistent with all observations, to be measured more precisely in future detection of gravity waves.  Their embedding into supergravity exploits  the hyperbolic geometry of the Poincar\'e disk or half-plane. These geometries are isometric under \Mobius transformations, which include the shift symmetry of the inflaton field. We introduce a new \Kahler potential frame that explicitly preserves this symmetry, enabling the inflaton to be light. Moreover, we include higher-order curvature deformations, which can stabilize a direction orthogonal to the inflationary trajectory. We illustrate this new framework by stabilizing the single superfield $\alpha$-attractors.   
\end{abstract}

\maketitle



\noindent
{\bf Introduction.}  Inflationary theory provides a simple explanation of the approximate homogeneity and isotropy of our world.  For a broad set of initial conditions, the solutions of the equations of motion for the inflaton field and the geometry of space rapidly approach an inflationary attractor solution which describes an exponentially expanding nearly uniform universe. Moreover, inflation provides a physical mechanism to generate the deviations from smoothness due to quantum fluctuations. CMB observations such as by Planck have tested and narrowed down the possibilities \cite{Ade:2015tva,Planck:2015xua}. In this paper we will discuss cosmological $\alpha$-attractors \cite{Kallosh:2013hoa,Kallosh:2013yoa,Cecotti:2014ipa}, which provide an excellent fit to the latest observational results for $\alpha \lesssim O(10)$, see figure 1. Similar to inflation itself, these attractors have the property that almost independent of the choice of the inflaton potential in these models, an inflationary universe comes out with the right value of the spectral index $n_s$ and a predicted tensor-to-scalar ratio $r$. 

Why do supergravity $\alpha$ attractors fit the data so naturally? The answer, as we clarify in this letter, is entirely geometric:  observational predictions of these models are to a large extent  determined by geometry of the moduli space, rather than by the potential. Hence one gains  a lot by formulating these models in a way that makes the symmetries of the moduli space manifest.  

We have already emphasized in \cite{Kallosh:2015zsa} that these models can be associated with the hyperbolic geometry of the Poincar\'e disk or half-plane, and the corresponding M\"{o}bius symmetry of the scalar kinetic terms. In this letter we will employ this geometry to serve and protect the {\em physical} symmetry that is relevant during inflation. Concretely, we will use the freedom in the choice of the \K\, frame in supergravity  to invoke a \K\, potential that is invariant under a distinct subgroup of the  M\"{o}bius group, protecting the shift symmetry of the \K\ potential in the inflaton direction.

\begin{figure}[t!]
\begin{center}
\includegraphics[width=8cm]{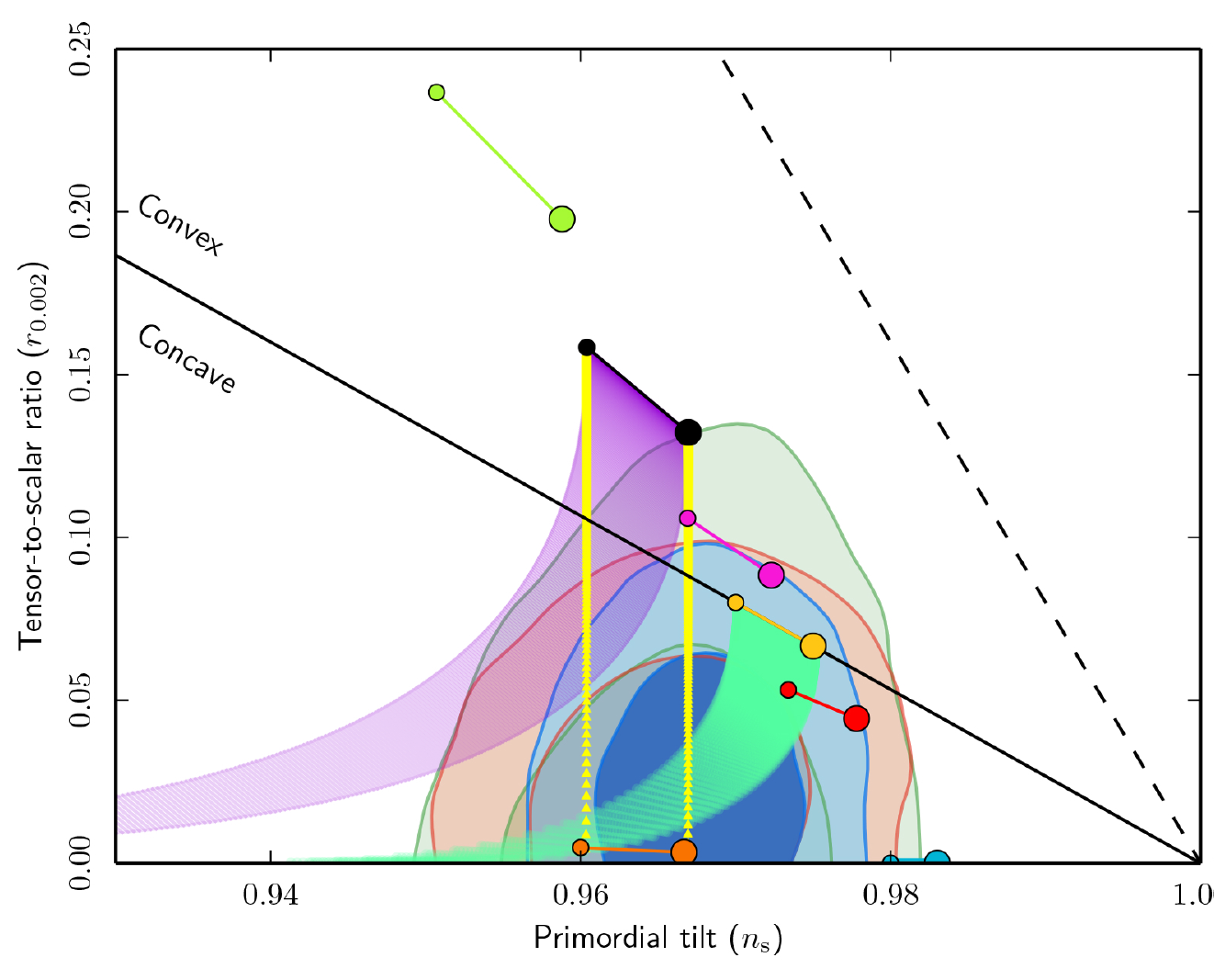}
\caption{\it The Planck/BICEP/Keck 2015 constraints on $n_s$ and $r$ with the predictions of a number of models \cite{Ade:2015tva,Planck:2015xua}. The yellow lines correspond to the simplest $\alpha$-attractor models for a full range $0<\alpha<\infty$ and $N=50,60$ \cite{Kallosh:2013yoa}. These predictions nicely fit the latest cosmological results for the most natural choice of $\alpha \lesssim O(10)$.}
\end{center}
\vspace{-0.5cm}
\end{figure}

The  original \K\, potential in half-plane variables is 
 \begin{align} \label{old-Kahler-H}
  K= -3\alpha \log (T+\overline T) \,.
\end{align}
The inflaton in these models resides in the real part of $T$, and the \K\, potential is not invariant under the shift of this field. 
We will  introduce a modified \K\, frame where the new \K\, potential  reads
\be \label{new-K}
 K = -{3\alpha\over 2} \log \left[{(T+\overline T)^2\over 4(T \overline T)}\right] \, .
\ee
It is related by a  \Kahler transformation to the original one. However, as we will show below, {\em it has a symmetry under the shift of the inflaton, accompanied by the rescaling of the inflaton partner}. This symmetry leads to the following feature of the new \K\, potential: during inflation, in these models the inflaton partner  $T-\overline T$ vanishes and $K=0$,  which is obviously invariant under the inflaton shift. This inflaton shift symmetry is only slightly broken by the superpotential, which now makes manifest a critical feature of $\alpha$-attractor  models:  {\em the inflaton is light.}

The situation is reminiscent of the mechanism of the inflaton shift symmetry proposed in \cite{Kawasaki:2000yn}.  There  the \K\, potential does not depend on the inflaton direction, which one can take as the real part of the chiral multiplet $\Phi$, and depends only on its partner:
$
  K = \tfrac12 (\Phi - \overline \Phi)^2 \,.
$
While this is related by a \Kahler transformation to the canonical case $K = \Phi \overline \Phi$, the former has a shift symmetry for the inflaton. Again, since only the superpotential breaks this symmetry of the  \K\, potential, the inflaton can be naturally light during inflation in the supergravity model of the quadratic chaotic inflation 
\cite{Kawasaki:2000yn}. Moreover, this construction can be generalized by including a generic function in  the superpotential. This results in a broad class of chaotic inflation model in supergravity with nearly arbitrary inflaton potentials proposed in \cite{Kallosh:2010xz}.

Our new \Kahler frame can be seen as the curved analogon of the flat \Kahler potential with a shift symmetry. In the limit $\alpha \rightarrow \infty$ where the curvature tends to zero, the new \Kahler potential \eqref{new-K} goes to $K = \tfrac12 (\Phi - \overline \Phi)^2$ after the identification $T = \exp( 2 \Phi / \sqrt{3 \alpha})$, as used in \cite{Roest:2015qya}. A peculiar property of both is that $K$ vanishes on the inflationary trajectory $\Phi = \overline \Phi$. 

\noindent
{\bf \Mobius transformations.} First we describe the necessary mathematical background.
The symmetry of the moduli space metric corresponds to the \Mobius group, both in disk and in half-plane variables. The metric in half-plane variables reads:
\be
 ds^2=  3\alpha {d T d \overline T \over (T+\overline T)^2} = 3 \alpha {d\tau d \overline \tau \over (2 \, {\rm Im} \tau)^2} \ ,
 \ee
where  $ \tau= i T$. The full set of isometries of this geometry can be generated by the following four transformations:
 \begin{itemize}
  \item Translation of the imaginary part: $T \rightarrow T - i b$,
  \item Dilatation of the entire plane: $T \rightarrow a^2 T$,
  \item Inversion : $T \rightarrow 1 /  T$,
    \item Reflection of the imaginary part: $T \rightarrow \overline T$.
 \end{itemize}
The three holomorphic combination of these, i.e.~translations, dilatations and  inversions, generate the following M\"{o}bius transformations:
\be
 \tau \rightarrow {a\tau+b\over c\tau +d},  \qquad \Delta \equiv ad-bc\neq 0 \,,
\label{sl}\ee
and $a,b,c,d$ are real numbers. The \Mobius group therefore corresponds to a transformation associated with an $GL(2, \mathbb{R})$  matrix
\begin{eqnarray}\label{mobius}
         \mathcal{M} =  \left(           \begin{array}{cc}
                      a&  b \\
                        c & d \\
                      \end{array}
                    \right) \in GL(2, \mathbb{R}) \,.
\end{eqnarray}
The Poincar\'e line~element above is invariant under any non-singular transformation. However, when restricting to a particular half-plane, this is only mapped onto itself when one takes the determinant $\Delta$ to be positive.

A general \Mobius transformation can be conveniently parametrized via the Iwasawa decomposition, 
 \begin{align}\label{KAN}
  \mathcal{M} & = K \cdot A \cdot N  \,, \notag \\
 & =  \left(           \begin{array}{cc}
                       \cos\theta &  - \sin\theta \\
                        \sin\theta & \cos\theta \\
                      \end{array}
                    \right) \cdot  
                    \left(           \begin{array}{cc}
                       r_1 & 0 \\
                       0 & r_2 \\
                      \end{array}
                    \right) \cdot  
                    \left(           \begin{array}{cc}
                       1 & x \\
                       0 & 1\\
                      \end{array}
                    \right) \,,
      \end{align}             
whose parameters are given by
 \begin{align}
  &  r_1 = \sqrt{a^2 + c^2} \,, \quad x = \frac{ab+cd}{a^2 + c^2} \,, \notag \\
  & r_2 = \frac{\Delta}{\sqrt{a^2 + c^2}}   \,, \quad \cos\theta = \frac{a}{\sqrt{a^2 + c^2}} \,.
 \end{align}
Here the $K \cdot A \cdot N$ subgroups parametrize the compact, Abelian and nilpotent transformations of the \Mobius group, respectively. In the case that $\Delta=1$ the symmetry is reduced to $SL(2, \mathbb{R})$.
 
\noindent
{\bf A new \K\ frame.} Now that we have phrased the Iwasawa decomposition, we  can  turn to explicit realizations of this geometry in terms of \Kahler potentials, and a discussion of the physical significance as to which of the isometries they preserve. 

 First let us address the expression of $T$ in canonical variables,
 \begin{align} \label{def-T}
   T = \exp\left( \sqrt{\frac{2}{3\alpha}} \, \varphi\right) + i \chi \,,
  \end{align}
where the dilatonic field $\varphi$ will be our inflaton, and $\chi$ our axion.  This physical realization is determined crucially by the geometry we choose to employ.
The $SL(2, \mathbb{R})$ symmetry of the kinetic terms of the axion-dilaton pair was first derived in the context of ${\cal N}=4$ supergravity in \cite{Cremmer:1977tt}. The nilpotent subgroup $N$  of the Iwasawa decomposition, relevant to the conventional \Kahler potentials, acts as a shift on the axionic field: 
 \begin{align} 
  \chi \rightarrow \chi + b \,.
  \end{align} 
In contrast, the Abelian dilatation shift symmetry, $A$, acts on both components, 
 \begin{align} \label{infl-shift}
   \chi \rightarrow \frac{a}{d} \chi \,, \quad 
   \varphi \rightarrow \varphi + \sqrt{\frac{3 \alpha}{2}} \log(a/d) \,.
   \end{align}
   Here we have presented a $GL(2, \mathbb{R})$ version of the symmetry with $\Delta \neq 1$.
Note that this acts as a shift symmetry on the field $\varphi$, which will play the role of the inflaton in our context.  In the case $\Delta=1$, as in $SL(2,\mathbb{R})$, the Abelian subgroup reads
 \begin{align}
   \chi \rightarrow  a^2 \chi \,, \quad 
   \varphi \rightarrow \varphi + \sqrt{{6 \alpha}} \log a  \,.
   \end{align} 
For infinitesimal transformations $a=1+\delta a$ this acts as
\begin{align}
\tanh  \Big (\frac{\varphi}{\sqrt{6 \alpha}}  \Big )  \rightarrow    \tanh \Big (  \frac{\varphi}{\sqrt{6 \alpha}} +\delta a \Big )  \,,
   \end{align}
on the inflaton field.

The conventional half-plane \Kahler potential is \eqref{old-Kahler-H}. This form of the potential is invariant under axionic symmetries that shift the imaginary component $\chi$, corresponding to the nilpotent subgroup $N$.  However, this property of the \K\, potential was not crucial in constructing axion inflation models.

The conventional Kahler potential is {\em not} invariant under the Abelian subgroup $A$ corresponding to dilatations. This group is particularly important as it corresponds to the shift symmetry of the inflaton, as we will show below. Therefore it would be valuable to highlight this shift symmetry in a \Kahler potential, and only introduce a (small) shift symmetry breaking via the superpotential. To this end we introduce a new \Kahler potential, which in half-plane coordinate is defined by
 \be \label{new-Kahler-H}
K_{\mathbb{H}}=  -3 \alpha \log \left( \frac{T + \overline T}{2 |T|}  \right) =-{3  \alpha\over 2}  \log \left( { \left(T + \overline{ T}    \right) ^2\over  4 \, T\, \overline{T}  } \right)  \,. 
\ee 
where $|T|= (T\, \overline{T})^{1/2}$. It is related to the old \Kahler potential by means of a \Kahler transformation. However, the symmetries of both potentials are different: the choices \eqref{old-Kahler-H} and \eqref{new-Kahler-H} are invariant under nilpotent and Abelian transformations, respectively, of which the latter correspond to the shift symmetry of the inflaton.

In detail, from the full set of \Mobius transformations \eqref{mobius}, the new \Kahler potential is preserved by the following transformations :
\begin{eqnarray}\label{dilatation}
 \mathcal{M} = \left(
                      \begin{array}{cc}
                      a&  0 \\
                        0 & d \\
                      \end{array}
                    \right) \,  : \qquad 
T\to {a T \over d}   \quad {\rm dilatation} \,, \notag \\
 \mathcal{M} = \left(
                      \begin{array}{cc}
                      0&  b \\
                       c & 0 \\
                      \end{array}
                    \right)  : \qquad 
T\to -{b\over c T}   \quad {\rm inversion} \ .
\end{eqnarray}
Note that in the Iwasawa decomposition the latter  can be seen as an arbitrary dilatation, $r_1=c$, $r_2=-b$, followed by a discrete inversion with $\theta=90$\textdegree \, in \rf{KAN}.  

\noindent
{\bf Disk variables.}  In disk variables, related to the half-plane variables by the Cayley transform,
\be
Z= {T-1\over T+1}\, ,    \qquad T= {1+Z\over 1-Z} \, ,
\label{Z1}
\ee 
the conventional \Kahler potential for  is
 \begin{align} \label{old-Kahler-D}
  K = -3 \alpha \log ( 1 - Z \overline Z ) \,, \quad 
  Z =  {e^{\sqrt{\frac{2}{3 \alpha}} \varphi} + i \chi-1\over e^{\sqrt{\frac{2}{3 \alpha}} \varphi} + i \chi+1} \,.
 \end{align}
Note that an interesting and cosmologically important feature of a disk variable is that at $\chi=0$ 
  \be
 Z|_{\chi=0}= \tanh \Big (\frac{\varphi}{\sqrt{6 \alpha}}  \Big )
 \ee
This \Kahler potential parametrizes the same geometry: it is related to the half-plane \Kahler potentials \eqref{old-Kahler-H} and \eqref{new-Kahler-H} by a Cayley and a \Kahler transformation. A consequence of the latter is that the symmetries of the \Kahler potential change. As is clear from the formulation in disk variables, this form of the \K\, potential has a rotational symmetry, corresponding to the compact group $K$. 

The new \Kahler potential reads
 \be \label{new-Kahler-D}
  K_{\mathbb{D}}= -{3  \alpha\over 2}  \log  \left[ {( 1- Z \overline{Z} )^2  \over  (1-Z^2) (1- \overline{Z}^2)}  \right] \,,
 \ee
which is related to $K_{\mathbb{H}}$ by means of a Cayley transformation, without the need of any supplementing \Kahler transformations. Therefore it has the same explicit symmetries. In disk variables, the dilatation symmetry acts as 
 \begin{align}
 ( 1 \pm Z) \rightarrow (1 \pm Z) \frac{{\beta} \pm {\gamma}}{{\beta} + Z {\gamma}}
 \end{align}
 with real parameters $\beta =(a+d)$ and $\gamma = (a-d)$. In addition, the inversion symmetry takes the form
  \begin{align}
  ( 1 \pm Z) \rightarrow (1 \mp Z) \frac{ \widetilde{\beta} \mp \widetilde{\gamma}}{\widetilde{\beta} + Z 
  \widetilde{\gamma}}
 \end{align}
again for real parameters $\tilde \beta = (b-c)$ and $\tilde \gamma = -(b+c)$. A particular case of this is $Z \rightarrow -Z$ and corresponds to a $\theta=90$\textdegree rotation in the Iwasawa decomposition.

Noting that dilatation takes
\begin{align}
( Z \overline Z - 1) \to \left(Z \overline{Z}-1\right) \frac{ ({\beta} -{\gamma} ) ({\beta} +{\gamma} )}{({\beta} +{\gamma}  Z)
   \left(\gamma  \overline{Z}+{\beta} \right)}\,, 
   \end{align}
it is apparent that the dilatation operation leaves invariant the quantity:
$$
{\cal I} =  \frac{Z \overline{Z} -1}{(1-Z)(1+\overline{Z})}\,.
$$
This object is also special under the inversion operation, which simply takes the conjugation, swapping between ${\cal I} \leftrightarrow \overline{\cal I}$.  
As such, the disk \Kahler potential makes  these symmetries manifest when the argument of the logarithm is written as $\mathcal{I} \mathcal{\overline I}$.

\noindent
{\bf Universal stabilization. } 
The crucial issue of the stabilization of the inflaton partner during inflation as well as the stabilization of both the inflaton and its partner at the minimum has been studied in detail over the years \cite{GomezReino:2006dk}, \cite{Achucarro:2007qa}. In particular,  in the case of a single superfield, the average mass of its two components reads \cite{GomezReino:2006dk, Louis} (see eqs.~(2.20) and (A.1) in \cite{Louis}): 
 \begin{align} \label{mass}
   m^2 = K^{\Phi\overline\Phi}\nabla_\Phi \nabla_{\overline\Phi} V=  3 (R + \tfrac23) m_{3/2}^2 + R \,  V \,, 
  \end{align}
for a critical point with $V'=0$, $DW\neq 0$, and where $R$ is the Ricci scalar. The above is only a condition  on the average of the two masses, of the inflaton and its partner,  and does not address stability in each of the two directions separately. Moreover, this condition is derived for $V' =0$. However, during inflation the potential is rather flat: the inflaton mass squared is much smaller than $V$ due to slow roll conditions. Therefore a positive average $m^{2}=\mathcal{O}(V)$ implies that the inflaton partner is stable during inflation.
Hence $R> -2/3$ is a necessary and  $R \geq 0$ is a sufficient condition for stabilization during inflation.

We will demonstrate that the crucial role played by the \Kahler curvature allows one to stabilize the inflaton partner in a universal way. To this end, we will  deform the maximally symmetric \Kahler manifold by adding quadratic and quartic terms: 
   \be
\hskip -0.1cm   K = \frac{ -3 \alpha}{1+2c_2} \log \Bigl( \frac{T + \overline T}{2 |T|} \Bigl[1+ c_2  \Bigl(\frac{T-\overline T}{T + \overline T}\Bigr)^{2}
    +c_{4} \Bigl(\frac{T-\overline T}{T + \overline T}\Bigr)^{4}\Bigr] \Bigr)  
 \label{stab}  \ee 
where we redefine the overall coefficient $3\alpha$ to include a set of $\alpha$-attractor \K\, potentials which have been studied in \cite{Cecotti:2014ipa, Kallosh:2015lwa}. Both of the new terms contribute to the curvature and can be used to stabilize the imaginary direction. 
Importantly, the higher-order terms preserve all symmetries of this \Kahler potential, both the dilatations and the inversions. We therefore retain the crucial inflaton shift symmetry, while breaking the axionic shift symmetry even stronger.

The curvature of the geometry based on \eqn{stab} is only constant in the maximally symmetric case with $c_2=c_4=0$. However, the curvature corrections are constant along the inflationary trajectory $T = \overline T$, leading to
  \begin{align}
    R = -\frac{2(1+8c_2+6c_2{}^2-12c_4)}{3 \alpha (1+2c_2)} \,.
  \end{align}
 Therefore the above conditions on stability during inflation, either the necessary one $R > -2/3$ or the sufficient one $R \geq 0$, can always be achieved, for any $\alpha$, by tuning $c_2$ and $c_4$.

Single-superfield models have seen a lot of progress recently; general potentials were constructed in \cite{Ketov:2014qha,Linde:2014ela}, one of the first inflationary models in supergravity was unearthed again \cite{Goncharov:1983mw,Linde:2014hfa} and the first examples of $\alpha$-attractors were constructed \cite{Roest:2015qya,Linde:2015uga}. The difference between the latter two resides in the choice of \Kahler potential. The first used the maximally symmetric case $c_2 = c_4 = 0$ leading to the following curvature and necessary stability condition \cite{Roest:2015qya}:
 \begin{align}
   R = - \frac{2}{3 \alpha} > - \frac{2}{3} \quad   \Leftrightarrow \quad \alpha > 1 \,.
 \end{align}
Instead, the second used the \Kahler potential introduced in \cite{Cecotti:2014ipa} with $2c_2 = 1- \alpha$ and $c_4=0$, leading to
  \begin{align}
 R = - \frac{2}{3 \alpha} -1 + \frac{1}{\alpha^2} > - \frac{2}{3} \quad   \Leftrightarrow \quad \alpha < 1 \,.
    \end{align}
As a consequence, their regimes of stability turn out to be complementary. From the above it follows that these stability constraints are mere consequences of the particular choice of higher-order terms. With the results of this letter, however, one can achieve stability for any configuration of $\alpha$ by including general quadratic and quartic terms \cite{WIP}.
 
A similar analysis of the stability cosmological attractors based  on two superfields, in case the second superfield is nilpotent \cite{Ferrara:2014kva}, has the following features. We consider models of the kind 
 \begin{align}
 K = K(\Phi , \overline \Phi,  S \overline S) \,, \quad W =  S f (\Phi) \,,
\end{align}
where $\Phi$ can be either half-plane or disk coordinates. Due to the nilpotency of $S$, we are only interested in stabilizing the inflaton partner during inflation. The average mass formula  is, again up to slow-roll corrections \cite{Kallosh:2010xz}:
 \begin{align}
m^{2} = & ( 1+ R_{\rm bs}  ) V, \quad  R_{\rm bs} = K^{\Phi \overline \Phi} K^{S \overline S} R_{\Phi \overline \Phi S\overline S}\, .
    \end{align}
 In this case it is therefore the bisectional curvature that determines stability. An example of this is provided by the  \Kahler potential 
 \begin{align}
 \hskip -0.2cm   K = -3 \alpha\log \left( \frac{T + \overline T}{2 |T|}   - { S \overline S\over {2 |T|}} 
 \Big [ 1-   c_{\rm bs} \Big( \frac{ T-\overline T }{T + \overline T} \Big)^2 \Big] \right), 
 \end{align} 
which leads to a bisectional curvature given by $ R_{\rm bs} = -(1 + 2 c_{\rm bs})/(3 \alpha)$.    Without the stabilization term this model is therefore unstable for $\alpha < 1/3$ \cite{Kallosh:2013yoa}.
However, it follows from this general discussion of the cosmological attractor models based  on two superfields (where the second superfield is nilpotent) that there is a universal geometric mechanism of stabilization of the inflaton partner during inflation, based on the bisectional curvature. The details will be described  separately in \cite{prep}.

\noindent
{\bf Conclusions. }
 We have proposed a new \Kahler potential for the hyperbolic geometry which preserves the shift symmetry of the inflaton $\varphi$. In terms of the Iwasawa decomposition into $K \cdot A \cdot N$ subgroups, the new \Kahler frame exactly picks out the relevant Abelian subgroup $A$ of the full \Mobius group. Higher-order corrections can be used to stabilize the orthogonal directions while retaining the inflaton shift symmetry. This improvement of the \K\ potential is especially adequate for investigation of cosmological attractors, which make cosmological predictions determined mostly by the geometry of the moduli space rather than by the details of the inflaton potential.

\noindent
{\bf Acknowledgements} JJMC, RK and AL are supported by the SITP and by the NSF Grant PHY-1316699. JJMC and RK are also supported by the Templeton foundation grant `Quantum Gravity Frontiers,' and AL is supported by the Templeton foundation grant `Inflation, the Multiverse, and Holography.'  DR would like to thank the SITP for its warm hospitality and stimulating atmosphere while this work was performed. 


\end{document}